\documentclass[aps,preprint,showpacs]{revtex4}
\usepackage{graphicx}
\usepackage{bm}

\newcommand{\vect}[1]{\mathbf{#1}}
\newcommand{\urho}{{\skew1\hat{\bm\rho}}}
\newcommand{\uphi}{{\skew1\hat{\bm\varphi}}}
\newcommand{\ui}{\mathbf{i}}
\newcommand{\uk}{\mathbf{k}}

\newcommand{\arcsinh}{\mathop{\mathrm{arcsinh}}\nolimits}

\begin{document} 

\title{Linear Momentum Density\\ in Quasistatic Electromagnetic Systems}
\author{
J M Aguirregabiria, 
A Hern\'{a}ndez 
and 
M Rivas 
} 
\affiliation{Theoretical Physics, 
The University of the Basque Country, \\
P.~O.~Box 644,
48080 Bilbao, Spain}

\bigskip

\begin{abstract} 
We discuss a couple of simple quasistatic electromagnetic systems
in which the density of electromagnetic linear momentum can be easily computed.
The examples are also used to illustrate
how the total electromagnetic linear momentum, which may also be
calculated by using the vector potential, can be understood as a 
consequence of the violation of the action-reaction principle, 
because a non-null external force is required to maintain constant 
the mechanical linear momentum.
We show how one can  avoid the divergence in the interaction linear
electromagnetic momentum of a system composed by an idealization often
used in textbooks (an infinite straight current) and a point charge.
\end{abstract} 

\pacs{PACS: 03.50.De}

\maketitle

\section{Introduction}\label{sec:intro}

The electromagnetic angular momentum of quasistatic systems has been
discussed in different examples in well known textbooks~\cite{feynman,heald}, as well
as in a number of articles~%
\cite{romer,pugh,corinaldesi,ah,lombardi,keyes,boos,romer2,bahder,ma,driver,sharma,%
griffiths,castro,johnson,chandler}. 
However, the electromagnetic
linear momentum appearing in the same kind of systems has comparatively attracted
less attention~\cite{calkin,casserberg,arh,butoli,arh2,johnson}, 
although it may be even more striking the first time a student is
faced with it. Furthermore, with the exception of the rather trivial
distribution in the example by Casserberg~\cite{casserberg} and the mathematically 
complex calculation in \cite{johnson},
in those articles only the total electromagnetic linear
momentum is computed by using its evolution equation or the vector potential.
The first goal of this work is to provide some simple
examples that can be used in the classroom to compute the density of
electromagnetic linear momentum, as described by the Poynting vector. 
In our first two examples, the total electromagnetic momentum can be computed 
both by integrating that density and directly by using the vector potential. 

In Poynting's theorem one is forced to assume  the electromagnetic field
carries linear momentum in order to get the right balance equation for
the total linear  momentum. This can be illustrated in the particular but
simplified context of our examples. In all of them the internal
electromagnetic forces will not satisfy the action-reaction principle,
so that to maintain constant the total mechanical momentum one has to
apply a non-null external force on the system.  Then, the  only way to
keep the balance equation (\emph{the total external force equals  the
derivative of the total linear momentum}) is to attribute a linear
momentum to the electromagnetic  field. In our examples this balance can
be simply checked and one may see how the electromagnetic momentum is
stored in the field while the system evolves starting from a
configuration in which some of its elements are separated at infinite
distance and thus with no electromagnetic momentum. All this may help
students to gain confidence on their understanding of the meaning of the
Poynting vector. 

In Section~\ref{sec:quasi} we review some properties of the kind of
quasistatic electromagnetic systems we will consider. Then, in
Sections~\ref{sec:torus} and~\ref{sec:line} we discuss two simple
systems in which one can easily compute the Poynting vector due to the
electric field of a point charge moving at constant small velocity and a
stationary magnetic field.  We explicitly check the balance equation of
the total linear momentum and also that it may be computed by using
either the vector potential or the Poynting  vector. We show that the
latter method, although more difficult in general, has the advantage of showing the
spatial distribution of momentum density; in fact, it uncovers a curious
fact about that distribution in the example of Section~\ref{sec:line}.
The total  electromagnetic linear momentum of the first model for the
system studied in that section is infinite. This divergence  illustrates
the limitations of an idealization often used in textbooks (an infinite
straight current) and can be avoided in several ways, one of which is
presented in the same section. In Section~\ref{sec:circuit} we analyze
another model which can be used to illustrate the relationship between
the violation of the action-reaction principle and the origin of the
electromagnetic momentum, as well as the origin of the divergence in the
first  model of Section~\ref{sec:line}. We also discuss
the energy balance in all the examples.

\section{Stationary current and point charge}\label{sec:quasi}

Let us consider an electromagnetic system formed by  a static magnetic
field $\vect{B}_I$ created by a stationary current $I$ 
as well as the electric field $\vect{E}_q$ and the 
magnetic field $\vect{B}_q$ created by a point charge $q$ moving with
a small constant velocity  (which may be null).  
In the total electromagnetic energy there will be a
constant magnetic energy associated to the density $B_I^2/2\mu_0$.
On the other hand, since the point charge, its fields and the self-contributions to the
energy density ($\epsilon_0E_q^2/2+B_q^2/2\mu_0$) and to the
linear momentum density ($\epsilon_0\, \vect{E}_q\times\vect{B}_q$)
move as a whole, the
associated electromagnetic energy and linear momentum are also constant (and infinite,
so that they
must be dealt with by means of some appropriate renormalization procedure).

But from now on we will forget about all these constant
self-contributions and consider only the interaction energy and
linear momentum densities given by the cross terms
$\vect{B}_q\cdot\vect{B}_I/\mu_0$ and $\epsilon_0\, \vect{E}_q\times\vect{B}_I$, 
respectively. 

The total interaction energy is then~\cite{energ}
\begin{equation}\label{eq:UqvA}
U=q\,\vect{\dot r}_q\cdot\vect{A}\left(\vect{r}_q\right),
\end{equation}
where the vector potential in the Coulomb gauge for $\vect{B}_I=\nabla\times\vect{A}$ is evaluated at the charge position $\vect{r}_q$.

Under the assumption of small constant velocity, the total interaction linear momentum
can be computed by using the result proved by Calkin~\cite{calkin} (see also~\cite{eyges}) for more general 
quasistatic systems,
which reduces in our case to
\begin{equation}\label{eq:PqA}
\vect{P}=q\mathbf{A}\left(\vect{r}_q\right).
\end{equation}
Calkin also proves in the more general context that the derivative of the 
electromagnetic linear momentum equals the external
force applied on the system to keep constant the mechanical linear momentum.
One of the goals of this work is to illustrate this kind of result
(as well as the general momentum balance~\cite{balance})
with some particularly easy examples; but we will compute not only the total linear
momentum but also its density, to gain some insight on its spatial distribution.

\section{Toroidal solenoid and point charge}\label{sec:torus}

\begin{figure}
\begin{center}
\includegraphics{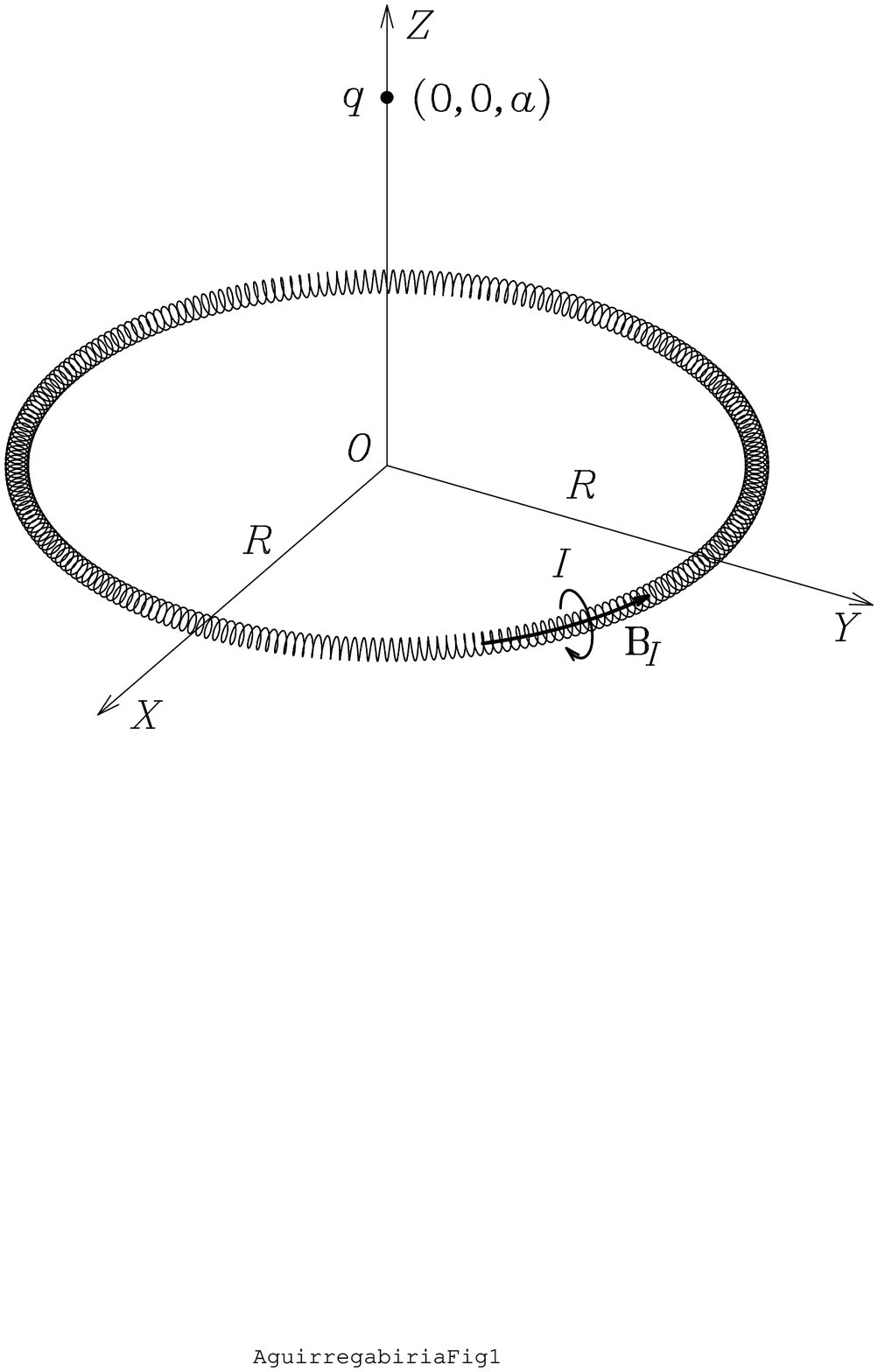}
\end{center}
\vspace{3cm}
\caption{Toroidal solenoid and point charge.\label{fig:torus}} 
\end{figure}

Our first example is the system depicted in Figure~\ref{fig:torus}:
a point charge is moving with a small constant velocity
along the symmetry axis
of a fixed toroidal solenoid. Since our goal is to illustrate in an example
a general result, we will make a twofold approximation which will
allow discussing the physics with a minimal mathematical apparatus. First
of all, we will assume that the torus is very thin so that the
area of the cross section is $S\ll R^2$, $R$ being solenoid's
main radius. With this assumption the magnetic field created by the current $I$
is confined inside the solenoid where its value, in the usual
cylindrical coordinates, is
 \begin{equation}\label{eq:bi}
\vect{B}_I=\frac{\mu_0NI}{2\pi R}\,\uphi,
 \end{equation}
assuming the solenoid has $N$ turns.
Furthermore, we will assume the point charge
located at $\vect{r}_q=a\,\uk$ moves slowly, so that we will
keep in all the calculations only the lowest order term in an expansion
in powers of $\beta\equiv\dot a/c$. In particular, we will check below that
this allows ignoring the change of the current
intensity on the solenoid due to the changing magnetic flux created by the charge,
because the induced electromotive force is proportional to $\beta^2$.
Moreover, this approximation reduces the exact expressions
for its electric and magnetic fields at a generic point of coordinates $(\rho,\varphi,z)$,~\cite{rete}
 \begin{eqnarray}\label{eq:Eqex}
 \vect{E}_q&=&\frac{\left(1-\beta^2\right)q}{4\pi\epsilon_0}\,\frac{\rho\,\urho+(z-a)\,\uk}{\left[\left(1-\beta^2\right)\rho^2+(z-a)^2\right]^{3/2}},
\\\label{eq:Bqex}
 \vect{B}_q&=&\frac1{c^2}\,\dot\vect{r}_q\times\vect{E}_q,
 \end{eqnarray}
to the easier
 \begin{eqnarray}\label{eq:Eq}
 \vect{E}_q&=&\frac{q}{4\pi\epsilon_0}\,\frac{\rho\,\urho+(z-a)\,\uk}{\left[\rho^2+(z-a)^2\right]^{3/2}},
 \\\label{eq:Bq}
 \vect{B}_q&=&\frac{\mu_0q\dot a}{4\pi}\,\frac{\rho}{\left[\rho^2+(z-a)^2\right]^{3/2}}\,\uphi.
 \end{eqnarray}

\subsection{Energy balance}

The total flux of the changing magnetic field $\vect{B}_q$
across the solenoid located at $({\rho=R, z=0})$ is
 \begin{equation}
 \Phi=N B_qS=\frac{\mu_0qNRS\dot a}{4\pi\left(R^2+a^2\right)^{3/2}},
 \end{equation}
so that the induced electromotive force is
 \begin{equation}
  \varepsilon=-\frac{d\Phi}{dt}=-\dot a\frac{d\Phi}{da}=
    \frac{3\mu_0qNRSa\dot a^2}{4\pi\left(R^2+a^2\right)^{5/2}}=
    \frac{3qNRSa}{4\pi\epsilon_0\left(R^2+a^2\right)^{5/2}}\,\frac{\dot a^2}{c^2}.
 \end{equation}
The dependence on $\beta^2$ of the last expression explicitly shows that, in our approximation, 
we could ignore induction
when computing the lowest order of the magnetic field created by the solenoid.
The power given to the solenoid is, then,
 \begin{equation}\label{eq:power}
  I\varepsilon=\frac{3\mu_0qNIRSa\dot a^2}{4\pi\left(R^2+a^2\right)^{5/2}}.
 \end{equation}

Since (in our low-velocity approximation) the solenoid creates no magnetic field outside the torus,
the point charge feels no magnetic force: no external force is necessary to keep its
constant velocity. But, due to the magnetic field created
by the charge, the solenoid will feel a magnetic force, so that an external force is necessary to keep it at rest. 
Since it is applied on a fixed solenoid, the external force performs no work.
Furthermore, the kinetic energy is constant.
So, where does power (\ref{eq:power}) come from? 

The answer is, of course, that the electromagnetic energy is not constant:
it has a time-dependent component due to the mixed electromagnetic field,
which can be computed by using (\ref{eq:UqvA}) or directly from
the density of electromagnetic energy:
 \begin{equation}\label{eq:elenergy}
 U=q\,\vect{\dot r}\cdot\vect{A}\left(\vect{r}_q\right)=\frac1{\mu_0}\int_T\vect{B}_q\cdot\vect{B}_I\,dV=
   \frac{RS}{\mu_0}\oint_T\vect{B}_I\cdot\vect{B}_q\,d\varphi=\frac{\mu_0qNIRS\dot a}{4\pi\left(R^2+a^2\right)^{3/2}}.
 \end{equation}
We can now see that  the derivative $\dot U$ is just the opposite of
the power given to the solenoid by the changing magnetic field $\vect{B}_q$,
so that the conservation of the total energy holds:
 \begin{equation}
  I\varepsilon+\dot U=0.
 \end{equation}

\subsection{Linear momentum balance}

As mentioned above, the point charge feels no magnetic force and there is no external force 
applied on it. However, we can easily check that the action-reaction principle is not
satisfied, for there is a magnetic force on the solenoid due to $\vect{B}_q$. 

Let us consider the current turns located between the angles $\varphi$ and $\varphi+d\varphi$.
We may think about them as an equivalent elementary turn of current
 \begin{equation}\label{eq:dI}
dI=NI\,\frac{d\varphi}{2\pi}, 
 \end{equation}
so that its magnetic dipole moment is
 \begin{equation}\label{eq:dm}
 d\vect{m}=dI\,S\,\uphi=\frac{NIS}{2\pi}\,d\varphi\,\uphi.
 \end{equation}
Incidentally, we can now check that the electromagnetic energy (\ref{eq:elenergy}) can also be computed by summing up the energies 
of these elementary dipoles:
\begin{equation}
U=\oint_Cd\vect{m}\cdot\vect{B}_q=\frac{\mu_0qNIRS\dot a}{4\pi\left(R^2+a^2\right)^{3/2}}.
\end{equation}

On the other hand, as a consequence of the magnetic field created by the moving charge, there is a force acting on the turn,
which can be easily calculated:
 \begin{eqnarray}\label{eq:dmdB}
 d\vect{m}\cdot\vect{B}_q&=&\frac{\mu_0qNIS\dot a}{8\pi^2}\,\frac{\rho}{\left[\rho^2+(z-a)^2\right]^{3/2}}\,d\varphi,\\
 d\vect{F}&=&-\left.\nabla\left(d\vect{m}\cdot\vect{B}_q\right)\right|_{\rho=R,\ z=0}
          =\frac{\mu_0qNIS\dot a}{8\pi^2}\,\frac{\left(2R^2-a^2\right)\urho+3Ra\,\uk}{\left(R^2+a^2\right)^{5/2}}\,d\varphi.
 \end{eqnarray}
The total magnetic force on the solenoid is now readily computed:
 \begin{equation}
  \vect{F}=\oint_Cd\vect{F}=\frac{3\mu_0qNISRa\dot a}{4\pi\left(R^2+a^2\right)^{5/2}}\,\uk.
 \end{equation}

Let us now assume that the particle came from infinity with a small constant velocity $\dot a <0$.
The particle feels no force, but to keep at rest the solenoid
one has to apply on it an external force $\vect{F}_{\mathrm ext}=-\vect{F}$.
The total impulse of this external force is readily computed:
 \begin{equation}\label{eq:impu}
 \int\vect{F}_{\mathrm ext}\,dt=\int_\infty^a\vect{F}_{\mathrm ext}\,\frac{da}{\dot a}=\int_a^\infty\frac1{\dot a}\,\vect{F} \,da=\frac{\mu_0qNIRS}{4\pi\left(R^2+a^2\right)^{3/2}}\,\uk.
 \end{equation}
At first sight one could think it is contradictory to have a net external impulse while
the mechanical linear momentum remains constant along the whole process. To solve
this puzzle and keep the general result that states \emph{the total impulse applied on a system
equals the change of its total linear momentum}, one has to take into account the linear momentum stored in the
electromagnetic field. 

The interaction electromagnetic linear momentum density in this setup
is confined inside the solenoid, where its value is
 \begin{equation}\label{eq:G}
\vect{G}=\epsilon_0\, \vect{E}_q\times\vect{B}_I=\frac{\mu_0qNI}{8\pi^2R}\,\frac{a\,\urho+R\,\uk}{\left(R^2+a^2\right)^{3/2}}.
 \end{equation}
We see that the linear momentum density has two  components which vanish at infinity. 
The one along the symmetry axis $OZ$ is always positive and
decreases monotonically from $a=0$ to $a\to\pm\infty$. The radial component has the sign of $a$
and reaches its maximum absolute value for $a=\pm R/\sqrt2$. 
By symmetry this radial component will give a total null contribution to the linear momentum.
In fact, the latter is easily computed, because in our approximation the integral extended to the inside 
of the torus $T$ reduces to an integral along its central line~$C$:
 \begin{equation}\label{eq:P}
\vect{P} =\int_T\vect{G}\,dV=SR\oint_C\vect{G}\,d\varphi=\frac{\mu_0qNIRS}{4\pi\left(R^2+a^2\right)^{3/2}}\,\uk.
 \end{equation}

The physical picture is now clear: the electromagnetic linear momentum (\ref{eq:P}) stored
inside the solenoid is a direct consequence of the impulse of the
external force $\vect{F}_{\mathrm ext}=-\vect{F}$ applied on the solenoid to
keep it at rest. If the particle came from infinity (where the electromagnetic linear momentum
was null) with a small constant velocity $\dot a <0$, the
total impulse of the external force equals the total electromagnetic momentum,
because from (\ref{eq:impu}) and (\ref{eq:P}) we have
 \begin{equation}
 \int\vect{F}_{\mathrm ext}\,dt=\vect{P}
 \end{equation}
or, equivalently, $\vect{F}_{\mathrm ext}\,dt=\dot{\vect{P}}$.

On the other hand, as a consequence of (\ref{eq:dm}), the vector potential created at charge's location by
the elementary turn $(\varphi,\varphi+d\varphi)$ is
 \begin{equation}\label{eq:dA}
d\vect{A}\left(\vect{r}_q\right)=\frac{\mu_0}{4\pi}\frac{d\vect{m}\times\left(\vect{r}_q-\vect{r}\right)}{\left|\vect{r}_q-\vect{r}\right|^3}
=\frac{\mu_0NIS}{8\pi^2}\,\frac{a\,\urho+R\,\uk}{\left(R^2+a^2\right)^{3/2}}\,d\varphi.
 \end{equation}
It is remarkable that, from the point of view of result (\ref{eq:PqA}), 
the linear momentum inside the elementary turn 
may be understood as created only by that turn, independently of the remaining turns in the solenoid:
 \begin{equation}
 d\vect{P}=\vect{G}\,dV=\vect{G}SR\,d\varphi=q\,d\vect{A}\left(\vect{r}_q\right).
 \end{equation}
Of course the same is not true outside the torus, where the linear momentum density is zero
due to the sum of contributions by all turns.

We can now check directly that (\ref{eq:PqA}) gives (\ref{eq:P}), because from the (\ref{eq:dA}) we find the total vector potential:
 \begin{equation}\label{eq:A}
\vect{A}\left(\vect{r}_q\right)=\oint_Cd\vect{A}\left(\vect{r}_q\right)=\frac{\mu_0NIRS}{4\pi\left(R^2+a^2\right)^{3/2}}\,\uk.
 \end{equation}
This direct method to compute the linear momentum is not
easier than the calculation involving the Poynting vector ---unless (\ref{eq:A}) is computed 
(by checking it gives the corretct $\vect{B}_I$, for instance) without calculating
(\ref{eq:dA})---, which would give us no hint on the
presence of a radial density of linear momentum.

\subsection{Angular momentum balance}

Since the dipole moment (\ref{eq:dm}) is parallel to the magnetic field created by the charge, there is no
torque on the turn, $d\vect{m}\times\vect{B}_q=0$, which is consistent with the constant
(null) value of the angular momentum. In fact, the mechanical angular momentum 
---with respect to the center of coordinates--- is null and 
the density of electromagnetic angular momentum is parallel to the solenoid,
 \begin{equation}\label{eq:rG}
\vect{r}\times\vect{G}=-\frac{\mu_0qNIR}{8\pi^2\left(R^2+a^2\right)^{3/2}} \,\uphi,
 \end{equation}
so that there is no electromagnetic angular momentum:
 \begin{equation}\label{eq:L}
\vect{L}=\int_T{\vect{r}\times\mathbf G}\,dV=SR\oint_C\vect{r}\times\vect{G}\,d\varphi=0.
 \end{equation}

\section{Straight current and point charge}\label{sec:line}

\begin{figure}
\begin{center}
\includegraphics{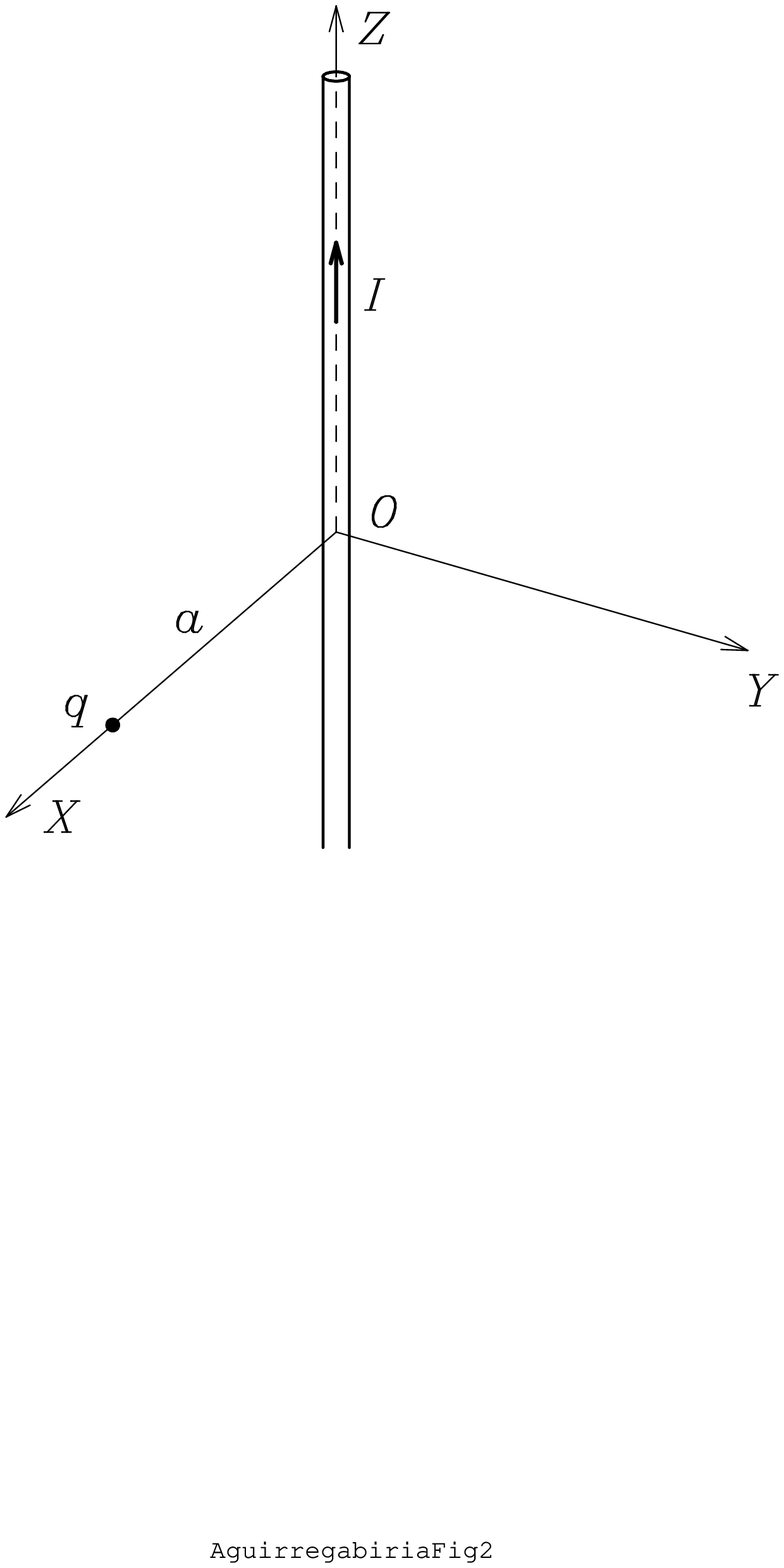}
\end{center}
\vspace{3cm}
\caption{Straight conducting wire and point charge.\label{fig:line}} 
\end{figure}

Let us now consider a fixed infinite straight wire conducting a constant current $I$
and a point charge directly moving away (or towards) the wire along a perpendicular 
direction with constant small velocity. If we choose cylindrical coordinates
around the $OZ$ axis of Figure~\ref{fig:line}, the charge position is given by
 \begin{equation}
\vect{r}_q=a\,\ui=a\left(\cos\varphi\,\urho-\sin\varphi\,\uphi\right). 
 \end{equation}
In the following we will neglect the wire radius and, 
as in the previous section, we will only keep at each step the lowest order in $\beta=\dot a/c$,
so that, at a generic point $\vect{r}=\rho\,\urho+z\,\uk$, 
the electric field of the charge and the magnetic field created by the current
are, respectively,
 \begin{eqnarray}\label{eq:wiree}
 \vect{E}_q&=&\frac{q}{4\pi\epsilon_0}\,\frac{(\rho-a\cos\varphi)\urho+a\sin\varphi\,\uphi+z\,\uk}%
 {\left(\rho^2+a^2+z^2-2\rho a\cos\varphi\right)^{3/2}},\\\label{eq:wirem}
 \vect{B}_I&=&\frac{\mu_0I}{2\pi\rho}\,\uphi.\label{eq:lBq}
 \end{eqnarray}

\subsection{External force}

By symmetry the magnetic field $\vect{B}_q$ created by the charge exerts no net force on
the wire, but, again, the action-reaction principle is violated
because there is a magnetic force exerted on the point charge by the magnetic field $\vect B_I$,
so that to keep constant the mechanical linear momentum one has to apply
on the point charge an external force opposite to the magnetic one:
\begin{equation}\label{eq:totextm}
 \vect{F}_{\mathrm ext}=-q\dot{\vect{r}}_q\times\vect{B}_I.
\end{equation}
Thus, we need a changing electromagnetic linear momentum whose
time derivative equals the total external force (\ref{eq:totextm}) applied on the system.

\subsection{Linear electromagnetic momentum}

\begin{figure}
\begin{center}
\includegraphics{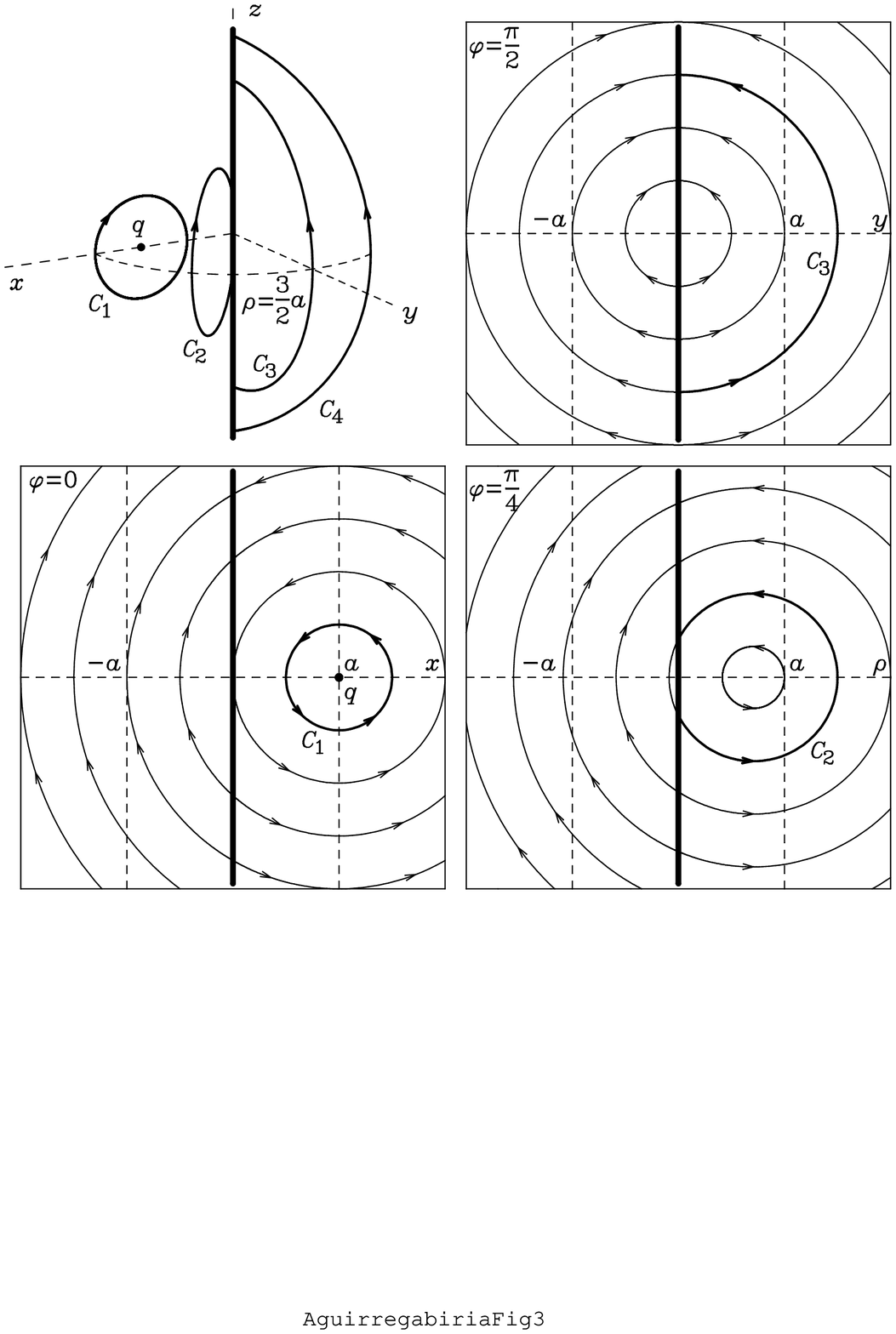}
\end{center}
\vspace{3cm}
\caption{Current lines of the linear momentum density (\ref{eq:gvect}).\label{fig:current}
In the upper left corner the lines passing through
$\displaystyle(\rho,\varphi,z)=(\frac{3a}2,\varphi,0)$ for $\displaystyle\varphi =0,\ \frac\pi4,\ \frac\pi2,\ \frac{3\pi}4$.
In the other three corners some lines in the planes $\displaystyle\varphi=0,\ \frac\pi4,\ \frac\pi2$ are
drawn, including ---in bolder line--- the one also displayed in the upper left corner.
The intersection between the plane and the cylinder $\rho=a$ appears as a couple of broken vertical lines.
}                                                            
\end{figure}

The cross term in the linear momentum density is easily computed from (\ref{eq:wiree})--(\ref{eq:wirem}):
 \begin{equation}\label{eq:gvect}
\vect{G}=\epsilon_0\, \vect{E}_q\times\vect{B}_I=\frac{\mu_0qI}{8\pi^2\rho}\,
  \frac{-z\,\urho+(\rho-a\cos\varphi)\,\uk}%
 {\left(\rho^2+a^2+z^2-2\rho a\cos\varphi\right)^{3/2}}.
 \end{equation}
Since $d\vect{r}\propto\vect{G}$ along the current lines of the linear momentum density,
their differential equations read as follows:
 \begin{equation}
\frac{d\rho}{-z}=\frac{dz}{\rho-a\cos\varphi},\qquad d\varphi=0.
 \end{equation}
The first equation is readily integrated to give $\left(\rho-a\cos\varphi\right)^2+z^2=C^2$, so
that the current lines of this density field are arcs of circumference
with center at $a\cos\varphi\,\urho$ lying in planes $\varphi=\mbox{const.}$
Some of them are depicted in Figure~\ref{fig:current}. By using
a well known  identity, we get
 \begin{equation}
 \nabla\cdot\vect{G}=\epsilon_0\,\vect{B}_I\cdot\left(\nabla\times\vect{E}_q\right)-
                     \epsilon_0\,\vect{E}_q\cdot\left(\nabla\times\vect{B}_I\right)=
                     -\frac1{c^2}\,\vect{E}_q\cdot\vect{j},
 \end{equation}
where $\vect{j}$ is the current density and we have used $\nabla\times\vect{E}_q=0$ and 
$\nabla\times\vect{B}_I=\mu_0\vect{j}$. So, as can also be proved
directly from (\ref{eq:gvect}), the divergence of $\vect{G}$
vanishes except inside the conducting wire, where it is positive for $z<0$ and negative for $z>0$.
 This in turn explains why its
current lines are closed or go from a wire point with $z<0$ to
another wire point with $z>0$.

Due to symmetry, the radial component of density (\ref{eq:gvect}) will vanish when integrated with respect to $z$,
so that the total electromagnetic linear momentum due to the interaction of the electric field
of the point charge and the magnetic field created by the current is 
 \begin{equation}\label{eq:Pwire}
\vect{P}=\int\vect{G} \,\rho\,d\rho d\varphi dz=\frac{\mu_0qI}{4\pi^2}\,\uk\,
     \int_0^\infty d\rho\int_0^{2\pi}d\varphi\,\frac{\rho-a\cos\varphi}{\rho^2+a^2-2\rho a\cos\varphi}
     =\frac{\mu_0qI}{2\pi}\,\uk\,\int_a^\infty \frac{d\rho}\rho.
 \end{equation}
We have used here the result (\ref{eqapi1}), which shows a curious fact: there is no net
contribution to the electromagnetic linear momentum for $\rho<a$, i.e., 
inside the cylinder centered at the current which 
goes through the charge. Although it is clear
from (\ref{eq:gvect}) ---and from Figure~\ref{fig:current}--- 
that this is only possible for $\rho<a$, because
only there does the sign of $G_z$ change when $\varphi$ varies, 
one might wonder whether this exact cancellation happens only
in the low velocity approximation we are using. The answer is that the same cancellation 
occurs when the
exact electric field of the point charge is used. This can be 
easily checked, because the linear momentum density in the relativistic case is
 \begin{equation}
\vect{G}=\frac{\mu_0qI\left(1-\beta^2\right)}{8\pi^2\rho}\,
  \frac{-z\,\urho+(\rho-a\cos\varphi)\uk}%
 {\left[\rho^2+a^2+\left(1-\beta^2\right)z^2-2\rho a\cos\varphi-\beta^2\rho^2\sin^2\varphi\right]^{3/2}},
 \end{equation}
so that using the integral (\ref{eq:I2}) we get 
 \begin{eqnarray}
\vect{P}&=&\int\vect{G} \,\rho\,d\rho d\varphi dz=\frac{\mu_0qI\sqrt{1-\beta^2}}{4\pi^2}\,\uk\,
     \int_0^\infty d\rho\int_0^{2\pi}d\varphi\,\frac{\rho-a\cos\varphi}{\rho^2+a^2-2\rho a\cos\varphi-\beta^2\rho^2\sin^2\varphi}\nonumber\\
     &=&\frac{\mu_0qI}{2\pi}\,\uk\,\int_a^\infty \frac{d\rho}\rho.\label{eq:Pwirerel}
 \end{eqnarray}

On the other hand, the contribution to the momentum (\ref{eq:Pwire}) for $\rho>a$ has a logarithmic divergence, as
one could expect from the slow decrease of the magnetic field (\ref{eq:lBq}) at infinity. 
This is just a consequence of the well known limitation of the model we are using: an infinite straight current 
is a good approximation to compute fields near the wire, but it may lead to meaningless
results for far fields, which enter along with near fields in the total value of the
electromagnetic momentum. This limitation can be easily avoided if we consider that the return current
goes along a coaxial conducting cylinder of radius $\rho=R>a$, so that it encloses both the wire and the particle. Then ${\vect{B}}_I$
and the Poynting vector will vanish outside the cylinder and instead of (\ref{eq:Pwire}) we have the finite
 \begin{equation}\label{eq:PwireR}
\vect{P}=\frac{\mu_0qI}{2\pi}\,\uk\,\int_a^R \frac{d\rho}\rho=\frac{\mu_0qI}{2\pi}\ln\frac Ra\,\uk.
 \end{equation}

The derivative of $\vect{P}$ is well defined in both models
(and directly related to the change due to the motion of the surface $\rho=a$ with separates
the regions with null and non-null contribution):
 \begin{equation}
 \dot{\vect{P}}=\dot a\,\frac{d\vect{P}}{da}= -\frac{\mu_0qI\dot a}{2\pi a}\,\uk=-q\dot{\vect{r}}_q\times\vect{B}_I
 =\vect{F}_\mathrm{ext}.
 \end{equation}
This is the right evolution equation for the linear momentum: 
the net external force applied on the system equals the time derivative of the total linear momentum.

The vector potential of the magnetic field created by the currents through the wire and the cylinder is
in our approximation
 \begin{equation}\label{eq:acyl}
 \vect{A}=
 \frac{\mu_0I}{2\pi}\,\theta(R-\rho)\,\ln\frac R\rho\,\uk,
 \end{equation}
where the Heaviside function is defined as usual:
 \begin{equation}\label{eq:Heaviside}
 \theta(x)\equiv\left\{\begin{array}{ll}
 1,&\mbox{if }x>0;\\
 0,&\mbox{if }x<0.
 \end{array}
 \right.
 \end{equation}
So, if we were interested only in the total electromagnetic linear momentum we could have used
(\ref{eq:acyl}) for $\rho=a$ in (\ref{eq:PqA}) to  compute (\ref{eq:PwireR}) more easily, but then we
would have known nothing about the curious distribution of its density.

In this example the magnetic force applied on the charge performs no work, so 
both the mechanical and the electromagnetic energies are constant. In fact, it is easy to see that
the density of electromagnetic energy $1/\mu_0\vect{B}_q\cdot\vect{B}_I$ is antisymmetric
with respect to the reflection $z\leftrightarrow-z$ and, in consequence, the total
electromagnetic energy due to the
interaction of $\vect{B}_q$ and $\vect{B}_I$ is zero, which is also a direct consequence of (\ref{eq:UqvA})
because the charge velocity and the vector potential are perpendicular.

By taking the limit $R\to\infty$ in (\ref{eq:PwireR}) we recover the divergence in (\ref{eq:Pwire}), 
but the latter can also be understood
by considering another family of models that recovers in the appropriate limit 
the one we have used at the beginning of this section.

\section{Circuit and point charge}\label{sec:circuit}

\begin{figure}
\begin{center}
\includegraphics{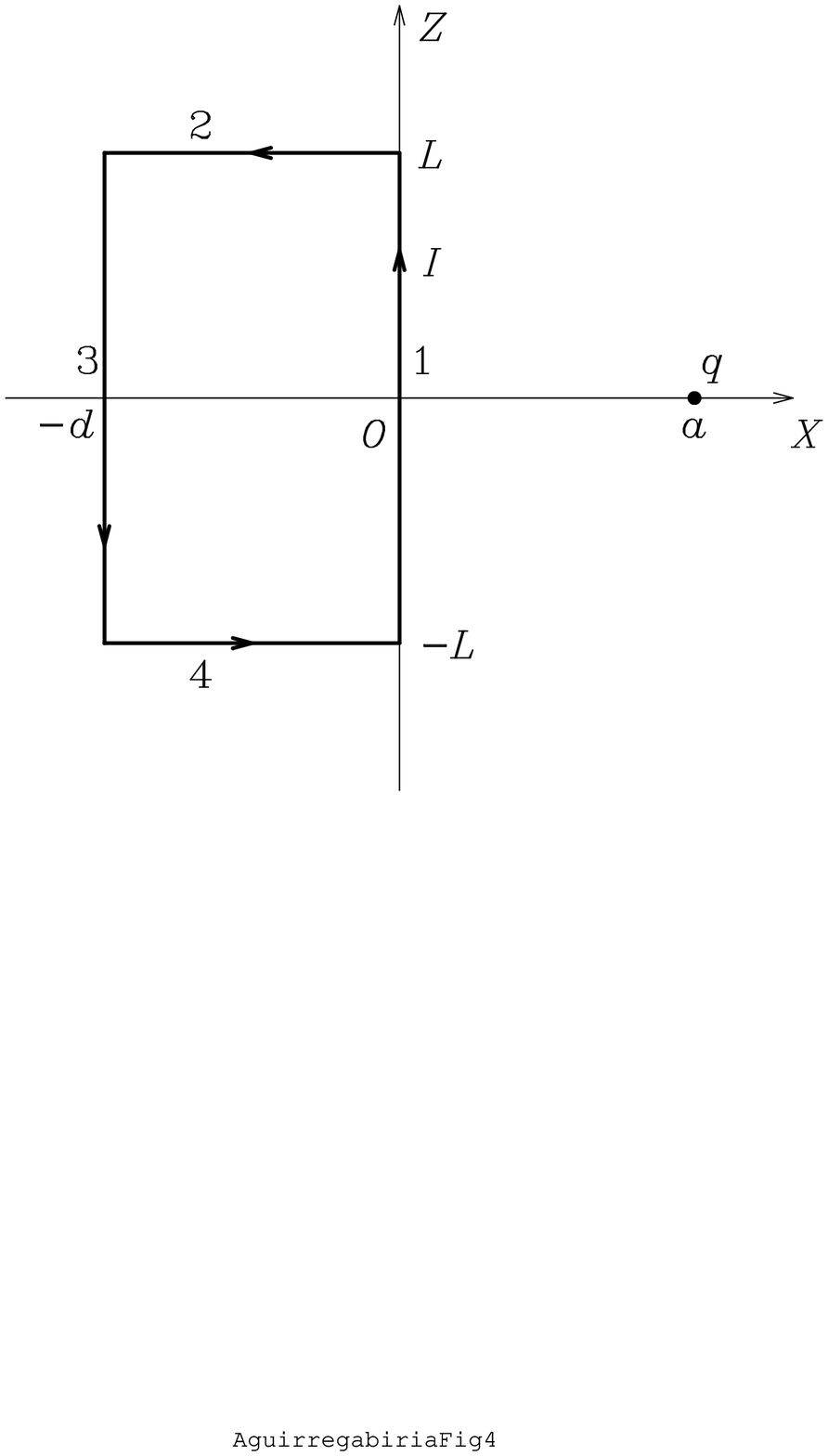}
\end{center}
\vspace{3cm}
\caption{Circuit and point charge.\label{fig:circuit}} 
\end{figure}

In this section we will consider the circuit of Figure~\ref{fig:circuit}
and the point charge moving with constant velocity along a symmetry axis.
If we take the double limit $L,\ d\to\infty$ we recover the system of Section~\ref{sec:line}.
By symmetry, the flux of the magnetic field created by the charge across the circuit is null,
so that there is no induction. 

In this case also the action-reaction principle is violated, for the
external forces that have to be applied on the charge (to keep constant
its velocity) and on the circuit (to keep it at rest) are not opposite
to each other. By symmetry, the magnetic force exerted on circuit sides 1 and 3
vanish, while those applied on sides 2 and 4 are equal, so that the total magnetic force
on the circuit is
 \begin{equation}
 \vect F_I=\frac{\mu_0qI\dot a}{2\pi}\left(\frac{a+d}{L\sqrt{(a+d)^2+L^2}}-\frac{a}{L\sqrt{a^2+L^2}}\right)\,\uk,
 \end{equation}
as one can easily check. It is not difficult to compute the magnetic field created by each
circuit side on the point charge. The force exerted by the magnetic fields
of arms 2 and 4  is just $-\vect F_I$, but there are also the forces due to the magnetic fields
created by 1 and 4: the total magnetic force on the charge happens to be
 \begin{equation}
 \vect F_q=\frac{\mu_0qI\dot a}{2\pi}\left(%
 \frac{L}{a\sqrt{a^2+L^2}}
 -\frac{L}{(a+d)\sqrt{(a+d)^2+L^2}}
 +\frac{a}{L\sqrt{a^2+L^2}}
 -\frac{a+d}{L\sqrt{(a+d)^2+L^2}}\right)\,\uk.
 \end{equation}
In consequence, to keep at rest the circuit and constant
the charge velocity, external forces must be applied on both elements, so that although
the mechanical linear momentum is constant, the net external force on the
system does not vanish:
 \begin{equation}
 \vect F_\mathrm{ext}=-\left(\vect F_I+\vect F_q\right)=
-\frac{\mu_0qI\dot a}{2\pi}\left(%
 \frac{L}{a\sqrt{a^2+L^2}}
 -\frac{L}{(a+d)\sqrt{(a+d)^2+L^2}}
\right)\,\uk.
  \end{equation}
Again we need a changing electromagnetic linear momentum to have the right evolution for the
total linear momentum.

The electromagnetic linear momentum due to the interaction
of fields created by circuit and charge can be easily computed by using~(\ref{eq:PqA}) and
 \begin{equation}\label{eq:Acir}
 \vect{A}\left(\vect{r}_q\right)=\frac{\mu_0I}{4\pi}\oint_C\frac{d\vect{r}}{\left|\vect{r}_q-\vect{r}\right|},
 \end{equation}
where $C$ is the circuit. By symmetry, the contribution from sides 2 and 4 to (\ref{eq:Acir})
are opposite: $\vect{A}_2\left(\vect{r}_q\right)+\vect{A}_4\left(\vect{r}_q\right)=0$. The contribution from side 3 is 
 \begin{equation}\label{eq:Acir3}
 \vect{A}_3\left(\vect{r}_q\right)=-\frac{\mu_0I}{4\pi}\,\uk\,\int_{-L}^L\frac{dz}{\sqrt{(a+d)^2+z^2}}= -\frac{\mu_0I}{2\pi}\,\arcsinh\frac L{|a+d|}\,\uk.
 \end{equation}
Since the contribution from the remaining side is $\vect{A}_1=-\vect{A}_3\Big|_{d\to0}$, the vector potential (\ref{eq:Acir})
at the location of the point charge is
 \begin{equation}\label{eq:Acirt}
\vect{A}\left(\vect{r}_q\right)=\frac{\mu_0I}{2\pi}\,\left(\arcsinh\frac L{|a|}-\arcsinh\frac L{|a+d|}\right)\,\uk.
 \end{equation}
Now we can check directly the evolution equation for the total linear momentum:
 \begin{equation}
 \dot{\vect P}=\frac d{dt}\left[q\vect{A}\left(\vect{r}_q\right)\right]=q\dot a\frac{d\vect{A}}{da}\left(\vect{r}_q\right)
 = \vect F_\mathrm{ext}.
 \end{equation}

On the other hand, from result (\ref{eq:Acirt}) it is easy to 
understand the origin of the logarithmic divergence of the vector potential and 
the linear momentum $q\vect{A}\left(\vect{r}_q\right)$ in the limit case of the beginning of Section~\ref{sec:line},
because if we compute in any order the double limit we get:
 \begin{equation}
\lim_{d\to\infty,\ L\to\infty}\vect{A}\left(\vect{r}_q\right)
 =\frac{\mu_0I}{2\pi}\,\uk\,\lim_{L\to\infty}\ln\frac{\sqrt{a^2+L^2}+L}{|a|} 
 =\frac{\mu_0I}{2\pi}\,\uk\,\lim_{d\to\infty}\ln\left|\frac{a+d}{a}\right|. 
 \end{equation}

As in Section~\ref{sec:line}, the
external forces perform no work and the charge velocity and the vector
potential  are perpendicular, so that the total electromagnetic energy
due to the interaction of $\vect{B}_q$ and $\vect{B}_I$ is zero.

\acknowledgments
This work was supported by The University of the Basque Country
(Research Grant~9/UPV00172.310-14456/2002).

\appendix
\section{A couple of useful integrals}

To compute (\ref{eq:Pwire}), we have used 
 \begin{equation}\label{eqapi1}
 I_1\equiv\int_0^{2\pi}\frac{\rho-a\cos\varphi}{\rho^2+a^2-2\rho a\cos\varphi}\,d\varphi=
 \frac{2\pi}\rho\,\theta(\rho-a)
 \end{equation}
and in  (\ref{eq:Pwirerel}) one needs the more general
 \begin{equation}\label{eq:I2}
 I_2\equiv\int_0^{2\pi}\frac{\rho-a\cos\varphi}{\rho^2+a^2-2a\rho\cos\varphi-\beta^2\rho^2\sin^2\varphi}\,d\varphi
 = \frac{2\pi}{\rho\sqrt{1-\beta^2}}\,\theta(\rho-a).
 \end{equation}
Integral (\ref{eqapi1}) ---as well as others appearing in the previous pages--- can be computed by using
tables~\cite{grad} or a computer algebra system, but both methods 
are far less helpful in order to calculate (\ref{eq:I2}).
In fact, the easiest way to prove results (\ref{eqapi1}) and (\ref{eq:I2}) 
is to consider in the complex plane the ellipse $C$ described
by
 \begin{equation}
 z=(\rho\cos\varphi-a)+i\rho\sqrt{1-\beta^2}\,\sin\varphi, \qquad\left(0\le\varphi<2\pi\right)
 \end{equation}
for constants $\rho>0$, $a>0$ and $0\le\beta<1$. 
Then, since the pole $z=0$ lies inside $C$ for $\rho>a$, we have 
 \begin{equation}\label{eq:I3}
I_3\equiv\mathrm{Im}\oint_C\frac{dz}z=\rho\sqrt{1-\beta^2}\, I_2=2\pi\theta(\rho-a),
 \end{equation}
which is equivalent to (\ref{eq:I2}) and reduces to (\ref{eqapi1}) for $\beta=0$.

\begin{figure}
\begin{center}
\includegraphics{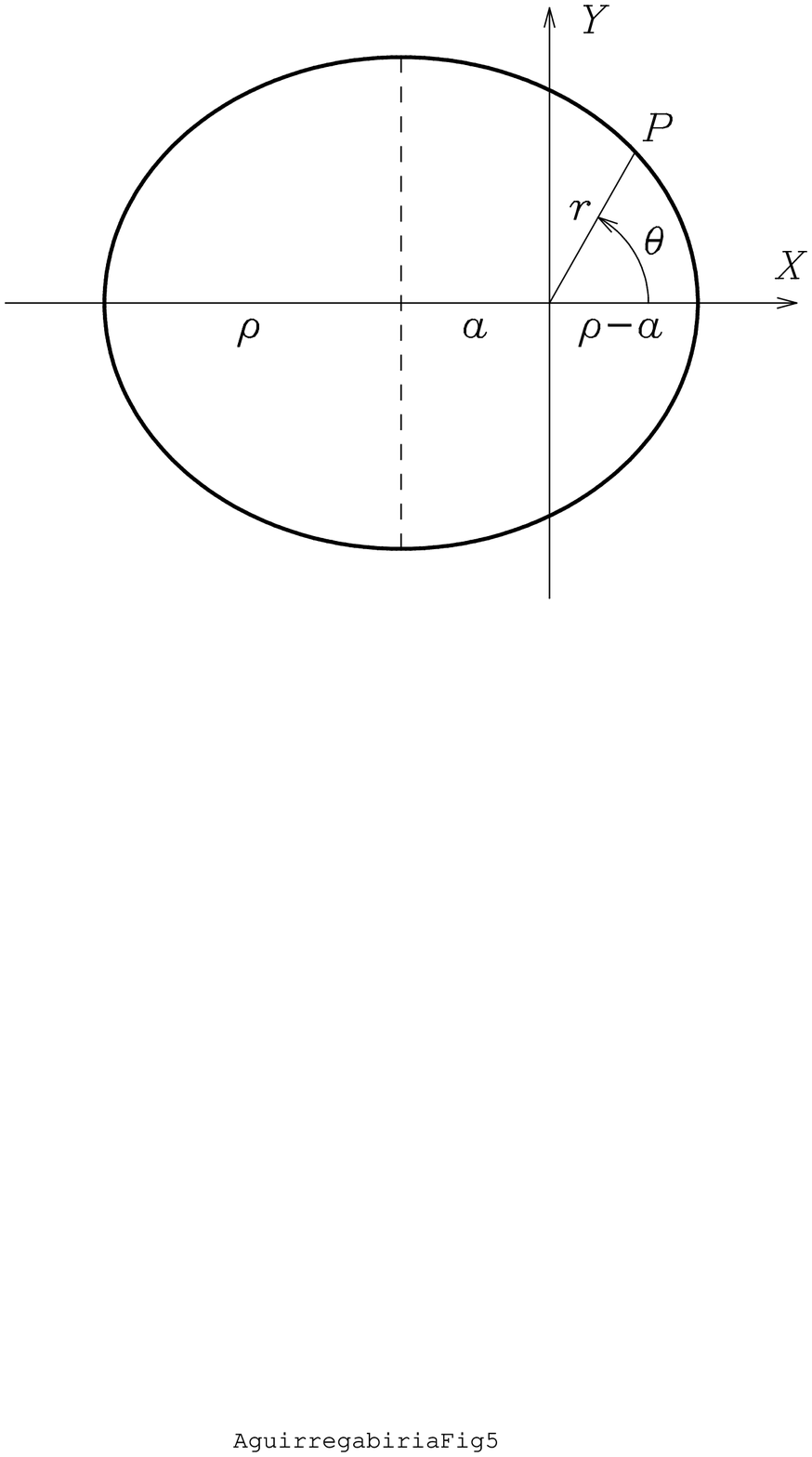}
\end{center}
\vspace{3cm}
\caption{Ellipse with center at $(-a,0)$ and semi-axes equal
to $\rho$ and $\rho\sqrt{1-\beta^2}$.\label{fig:ellipse}} 
\end{figure}

For students who do not know about complex integration, one might use
an equivalent but more pictorial method. By eliminating the parameter $0\le\varphi<2\pi$,
it is easy to check that the equations
 \begin{equation}
 x=\rho\cos\varphi-a,\qquad y= \rho\sqrt{1-\beta^2}\,\sin\varphi
 \end{equation}
describe an  ellipse with center at $(-a,0)$, horizontal major semi-axis of length
$\rho$ and eccentricity $\beta$, as depicted in Figure~\ref{fig:ellipse}. If we
consider the polar coordinates $(r,\theta)$ of an ellipse point~$P$, the integral (\ref{eq:I3}) is just
the full change of the polar angle $\theta$ in a complete counterclockwise 
turn of $P$ around the ellipse:
 \begin{equation}
\Delta\theta= \oint_C d\theta=\oint_Cd\arctan\frac yx=\oint_Cd\arctan\frac{\rho\sqrt{1-\beta^2}\,\sin\varphi}{\rho\cos\varphi-a}=\rho\sqrt{1-\beta^2}\, I_2.
 \end{equation}
It is not difficult to convince oneself that $\Delta\theta$ equals $2\pi$ if the origin is inside the
ellipse (i.e., if $\rho>a$) and $0$ otherwise. 



\clearpage
\section*{Figure captions}
\begin{description}
\item[Fig 1] Toroidal solenoid and point charge

\item[Fig 2] Straight conducting wire and point charge

\item[Fig 3] Current lines of the linear momentum density (\ref{eq:gvect}).
In the upper left corner the lines passing through
$\displaystyle(\rho,\varphi,z)=(\frac{3a}2,\varphi,0)$ for $\displaystyle\varphi =0,\ \frac\pi4,\ \frac\pi2,\ \frac{3\pi}4$.
In the other three corners some lines in the planes $\displaystyle\varphi=0,\ \frac\pi4,\ \frac\pi2$ are
drawn, including ---in bolder line--- the one also displayed in the upper left corner.
The intersection between the plane and the cylinder $\rho=a$ appears as a couple of broken vertical lines.
                                                            
\item[Fig 4] Circuit and point charge

\item[Fig 5] Ellipse with center at $(-a,0)$ and semi-axes equal
to $\rho$ and $\rho\sqrt{1-\beta^2}$
\end{description}


\clearpage

\clearpage

\clearpage

\clearpage

\clearpage

\end{document}